# Mechanics of freely-suspended ultrathin layered materials

*Andres Castellanos-Gomez*[1,2,*], *Vibhor Singh*[1], *Herre S.J. van der Zant*[1] and *Gary A. Steele*[1]

[1] Kavli Institute of Nanoscience, Delft University of Technology, Lorentzweg 1, 2628 CJ Delft (The Netherlands).

[2] Present address: Instituto Madrileño de Estudios Avanzados en Nanociencia (IMDEA-Nanociencia), 28049 Madrid (Spain).

* Email: a.castellanosgomez@tudelft.nl or andres.castellanos@imdea.org

## ABSTRACT

The study of atomically thin two-dimensional materials is a young and rapidly growing field. In the past years, a great advance in the study of the remarkable electrical and optical properties of 2D materials fabricated by exfoliation of bulk layered materials has been achieved. Due to the extraordinary mechanical properties of these atomically thin materials, they also hold a great promise for future applications such as flexible electronics. For example, this family of materials can sustain very large deformations without breaking. Due to the combination of small dimensions, high Young's modulus and high crystallinity of 2D materials, they have attracted the attention of the field of nanomechanical systems as high frequency and high quality factor resonators. In this article, we review experiments on static and dynamic response of 2D materials. We provide an overview and comparison of the mechanics of different materials, and highlight the unique properties of these thin crystalline layers. We conclude with an outlook of the mechanics of 2D materials and future research directions such as the coupling of the mechanical deformation to their electronic structure.

# 1. Introduction

The recent isolation of atomically thin materials by exfoliation of layered materials has opened a vast new field in the materials science [1-5]. These novel 2D materials have attracted attention of the scientific community because of their electrical, optical and magnetic properties (which usually differ from those of their bulk counterparts) but, moreover, they have also shown outstanding mechanical properties unmatched by conventional 3D materials. For example, single-layer graphene has an ultrahigh Young's modulus of 1TPa and it can sustain strains up to 25% without breaking [6].

In the past years, the scientific community has expanded beyond the electronic and optical properties to explore the mechanics of 2D layered materials. Questions such as how atomically thin materials respond to mechanical strain on a nanoscale have been studied in static deformation experiments. The 2D materials have been shown to have large Young's modulus, low residual stress and spectacularly large breaking strength. The mechanics of freely suspended 2D materials have also been explored in dynamic experiments in which the 2D materials were used as mechanical resonators [7]. Due to their light mass, such resonators are attractive for many applications such mass sensing [8]. By studying the mechanical resonator modes, phenomena such as plate-like to membrane-like transition can be observed [9, 10]. Dynamics also give access to characterize damping/energy loss mechanisms, which for 2D materials have been well characterized but are not yet fully understood [11, 12].

In the final section of this review, we provide an outlook towards future exciting directions with these freely suspended 2D materials, such as coupling of the mechanical deformation to the electronic structure. These effects that can be extraordinarily strong in 2D materials, providing a route towards hybrid optoelectronic-mechanic devices [13, 14].

## 2. Isolation and fabrication of 2D materials

Different fabrication methods have been developed to isolate atomically thin 2D materials in the past years [15]. The selection of the fabrication/isolation method strongly depends on the application as different techniques may yield different size, thickness and quality of the fabricated 2D crystals and hence different electrical and mechanical properties. In the next sub-sections the main routes to isolate 2D materials will be highlighted.

### 2.1. Mechanical exfoliation

Mechanical exfoliation (also referred to in the literature as micromechanical cleavage) has proven to be a simple yet powerful technique to obtain high-quality two-dimensional sheets by repeatedly cleaving bulk layered materials [16]. The weak van der Waals interaction between the layers makes it possible to cleave thin crystalline flakes by peeling off the surface of a bulk layered material that is adhered to a piece of sticky tape. These crystallites can be transferred to an arbitrary substrate by gently pressing the tape against the surface of the substrate and peeling it off slowly. More details on this technique can be found in the pioneering work of Novoselov, Geim *et al*. [17]

Although this method can produce high-quality flakes, its main drawback is the lack of control on the deposition step as flakes with various thicknesses are transferred all over the surface and only a small fraction are a few-layers thick. Nevertheless, it has been shown that one can identify 2D flakes, and distinguish them from thicker counterparts, in a fast and reliably way by optical microscopy. Important aspect of this technique is that atomically thin crystals show a characteristic color when they are deposited on top of certain substrates, due to a combination of interference color and optical absorption, which depends on the number of layers (see Figure 1a). There are several works that provide protocols to optimize the optical contrast and to optically determine the number of layers of many 2D materials:

graphene [18-22], graphene oxide [23], $MoS_2$ [24-26], $NbSe_2$ [24, 27], $WSe_2$ [26, 27], $TaS_2$ [26], $TaSe_2$ [28], mica [29, 30], etc.

As an example, Figure 1a shows an optical image of a typical $MoS_2$ flake obtained by mechanical exfoliation, containing a single-layer (1L) and a bilayer (2L) region. Although mechanically exfoliated graphene flakes can reach areas up to ~1 $mm^2$ in exceptional cases, most of the 2D materials fabricated by mechanical exfoliation are typically 10 to 1000 $\mu m^2$ in area.

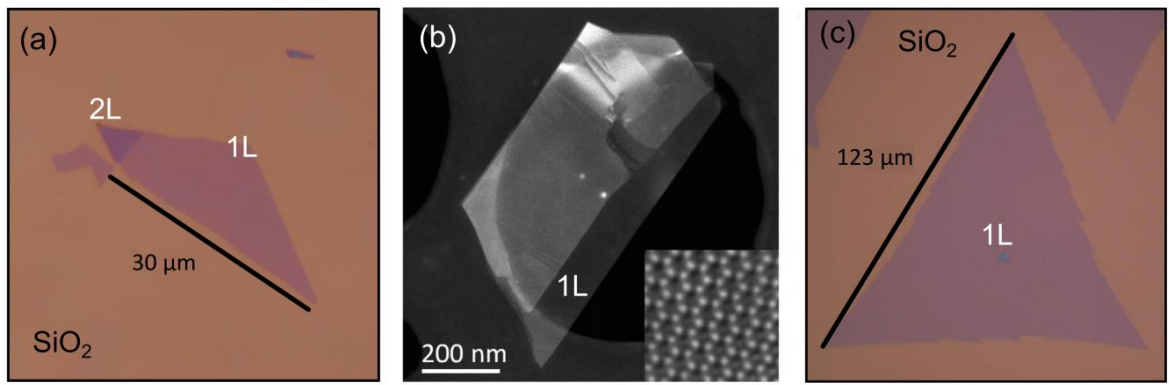

**Figure 1. Isolation of single-layer $MoS_2$ by different techniques.** (a) Optical image of a $MoS_2$ flake deposited on a $SiO_2$/Si substrate by mechanical exfoliation. (b) Transmission electron microscopy image of a $MoS_2$ flake isolated by ultrasonication of $MoS_2$ powder in an organic solvent (adapted from [31]). The inset shows a high-resolution scanning transmission electron microscopy image of the crystal lattice of the single layer region. (c) Optical image of a single-layer $MoS_2$ crystal grown on a $SiO_2$/Si by chemical vapour deposition method (adapted from [32]).

## 2.2. Liquid-phase exfoliation

Alternatively to the mechanical exfoliation technique, one can also isolate atomically thin 2D crystals by exfoliating bulk layered materials immersed in a liquid medium. Two main approaches are typically employed: exfoliation by direct sonication in a solvent or by employing an intercalant chemical agent and subsequent expansion. Details on the first approach can be found in Refs. [31, 33, 34] and we address the readers to Refs. [35-39] for details regarding the second liquid-phase exfoliation approach.

Due to the combination of high yield and low cost, this technique is a prospective fabrication approach to fabricate large quantities of atomically thin crystals. Liquid-phase exfoliation can yield suspensions of flakes with thicknesses ranging from 1 to 10 layers and areas of about 0.2-1 $\mu m^2$. The main limitations of this technique are the small size of the obtained flakes (too small for many applications) and the lack of sufficient control on the thickness to attain a monodisperse solution. Note that most of the studied nanomechanical devices have larger lateral sizes than the typical chemically exfoliated flakes. Figure 1b shows a transmission mode electron microscopy image of a $MoS_2$ flake fabricated by sonication of a $MoS_2$ powder [31]. The inset in Figure 1b shows the crystal lattice of the single-layer $MoS_2$ flake, acquired on the freely suspended part.

## 2.3. Chemical vapour deposition or epitaxial growth

Previously introduced methods to isolate 2D materials are top-down approaches, *i.e.*, 2D individual sheets are extracted from a parent 3D layered material. One can, on the other hand, also fabricate 2D materials by using recently developed bottom-up approaches where 2D layers are synthesized by assembling their constituent elements by thermal processing. Bottom-up synthesis of graphene can be carried out by epitaxial growth on silicon carbide substrates [40], by chemical vapour deposition (CVD) on copper [41] or nickel [42] substrates, by carbon segregation or by thermal decomposition of organic molecules on the surface of transition metals [43-46]. Note that graphene grown on metal surfaces usually needs a transfer step to fabricate graphene-based devices on standard $SiO_2$/Si substrates.

Other 2D materials have also been synthesized by bottom-up approaches such as vapour transport method [47], chemical vapour deposition [48, 49] or van der Waals epitaxy [50]. High-quality molybdenum and tungsten dichalcogenides, for instance, can be directly grown on $SiO_2$/Si substrates with single-crystal domains of 10 to 10000 $\mu m^2$ [32, 51, 52]. Figure 1c shows an optical microscopy image of a typical single-crystalline $MoS_2$ monolayer grown by CVD [32].

# 3. Fabrication of freely-suspended 2D materials for nanomechanical studies

The study of the mechanical properties of atomically thin materials typically requires the fabrication of freely-suspended samples such as doubly-clamped beams or circular drums. Three main approaches are employed to fabricate these suspended structures: direct exfoliation (flakes randomly distributed) onto pre-patterned substrates with holes/trenches [6, 10, 53-67], etching the substrate underneath the flakes [8, 12, 68-73] or depositing the flakes directly onto a specific hole or trench in the substrate using a transfer technique [9, 11, 74-89].

Figure 2a shows an optical image of a bi- and monolayer graphene flake that has been exfoliated by the conventional mechanical exfoliation method with an adhesive tape on a $SiO_2$/Si substrate pre-patterned with circular holes [67]. The graphene flake covers the holes forming 'micro-drumheads'. Figure 2b shows high-angle SEM images of freely-suspended monolayer graphene devices fabricated by etching the substrate underneath the graphene [8, 69]. Note that when a wet etching process is employed to remove the substrate, special care has to be taken to avoid the collapse of the suspended flakes by the capillary forces. Several examples of suspended devices fabricated by various transfer techniques are presented in Figure 2c. Mechanically exfoliated flakes or CVD grown layers can be transferred onto holes or trenches at specific locations on the substrate to make suspended devices [15, 90, 91]. Some recently developed techniques also allows one to transfer 2D materials already clamped to a pre-fabricated structure to avoid its collapse and/or to make electrical contact to the 2D layer (see the images in the central panel in Figure 2c) [80, 84, 88].

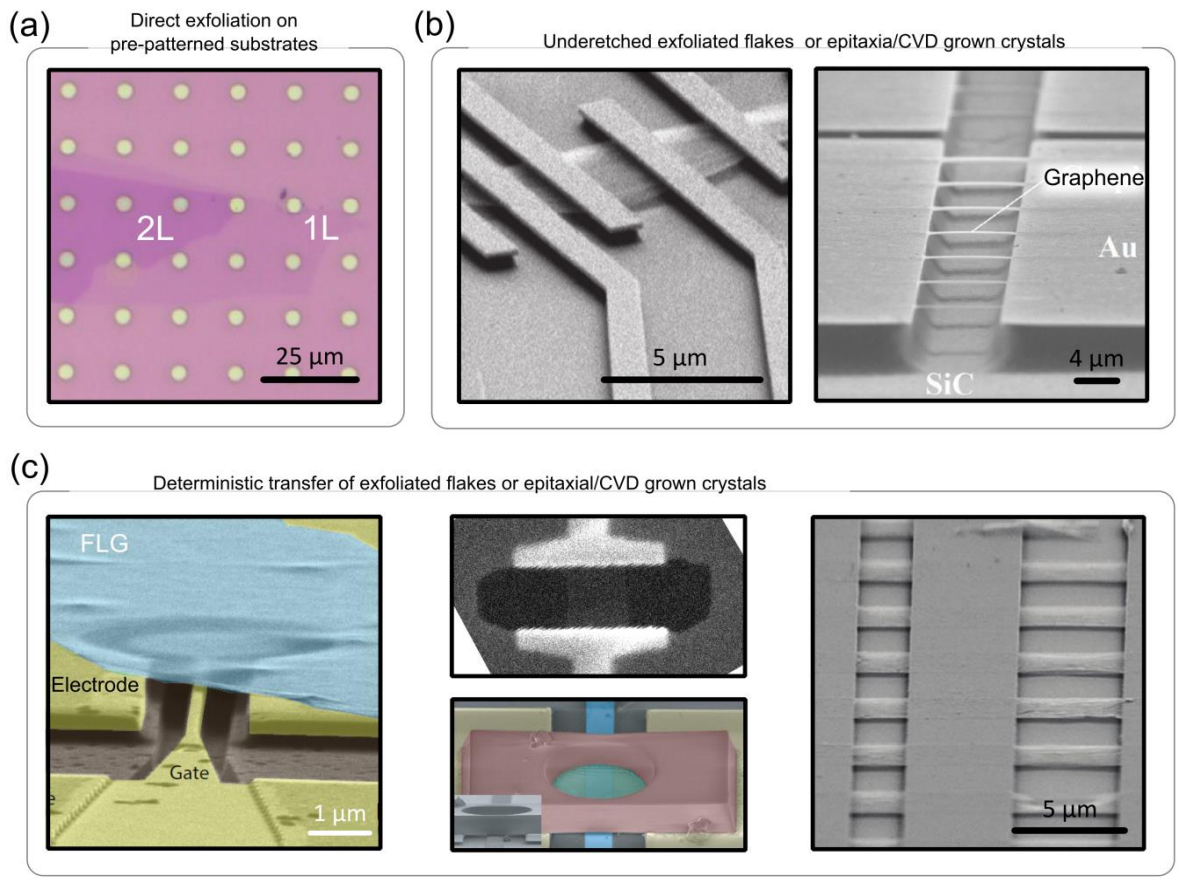

**Figure 2. Fabrication of freely suspended structures based on 2D materials.** (a) Optical image of a graphene flake exfoliated directly onto a surface pre-patterned with holes (adapted from [67]). (b) High angle scanning electron microscopy (SEM) images of suspended

monolayer graphene devices fabricated by etching the substrate underneath the graphene (adapted from [8] and [33]). (c) SEM images of different examples of graphene-based mechanical resonators, fabricated by transferring exfoliated or CVD grown graphene (adapted from [87], [80], [84] and [77]).

## 4. Static mechanical properties of suspended 2D materials

The mechanical properties of atomically thin materials have been extensively studied through the analysis of force *vs*. deformation experiments. In these experiments, a force load is applied to a freely suspended 2D crystal while its deformation is recorded. The following sub-sections describe the different experimental approaches, developed in the past years to characterize the static mechanical properties of nanolayers.

### 4.1. Central indentation experiment

One of the most employed approaches to study the elastic properties and breaking strength of freely suspended 2D materials is based on the analysis of the force *vs*. deformation traces acquired at the center of the suspended drum or doubly-clamped beam (see the inset in Figure 3a). In these experiments, an atomic force microscope (AFM) or a nanoindenter is used to apply a point load at the center of the suspended layer and to measure the subsequent deformation (the deflection of the suspended layer right at the center where the load is applied). For doubly-clamped structures, the force ($F$) *vs*. deformation ($\delta$) relationship follows the expression [54, 61]:

$$F = \left[ \frac{30.78\, W\, t^3}{L^3}\, E + \frac{12.32}{L}\, T \right] \delta + \frac{8\, W\, t\, E}{3\, L^3}\, \delta^3 \quad , \qquad [1]$$

where $W$, $L$, and $t$ are the width, length and thickness of the beam, $E$ is the Young's modulus and $T$ is the initial pre-tension. Note that the initial pre-tension values of freely suspended 2D materials are typically small: 0.07-1 Nm$^{-1}$ for graphene [6, 54, 57], 0.04-0.07 Nm$^{-1}$ for graphene oxide [76], 0.02-0.2 Nm$^{-1}$ for MoS$_2$ [65, 78] and 0.06-0.2 Nm$^{-1}$ for mica [63]. Freely suspended 2D materials show smaller pre-stress values (~0.01-1 GPa) compared to 10 to 200 nm thick silicon nitride nanomembranes (1-10 GPa). Despite the small tension value, single- and bilayer 2D materials are still in the membrane limit because of the vanishing bending rigidity for those thicknesses.

For a circular, drum-like structure, the force *vs*. deformation relationship follows [6, 65, 78]:

$$F = \left[\frac{4\pi t^3}{3(1-\nu^2)R^2}E + \pi T\right]\delta + \frac{tE}{(1.05-0.15\nu-0.16\nu^2)^3 R^2}\delta^3 \qquad , \qquad [2]$$

where ν is the Poisson's ratio and $R$ is the radius of the drum. The linear part in Eq. (1) and (2) includes a term that accounts for the bending rigidity of the layer (the first term, proportional to the Young's modulus) and second term that accounts for the initial pre-tension (proportional to $T$). Therefore, the analysis of the linear term of $F$ *vs*. $\delta$ traces does not allow one to unambiguously determine the Young's modulus and pre-tension of the suspended flakes. The cubic term in (1) and (2), that accounts for the stiffening in the layer due to the tension induced by the deflection, only depends on the Young's modulus and geometrical factors. Therefore, by fitting experimental non-linear $F$ *vs*. $\delta$ traces to (1) or (2) (depending on the geometry of the suspended nanostructure) one can determine the Young's modulus and the pre-tension independently from each other. This method to obtain the Young's modulus and pre-tension values typically yields 20-30 GPa and 0.02-0.03 N/m of uncertainty respectively.

According to expressions (1) and (2), for very thin layers the first term (bending rigidity) being proportional to $t^3$ may become negligible in comparison to the pre-tension and the deformation-induced tension terms. This is the case of single- and bilayer graphene, $MoS_2$, etc. in which the $F$ *vs.* $\delta$ traces are always highly nonlinear, allowing them to be modelled neglecting the bending rigidity term in Eq. (1) or (2) [6, 66, 78, 83, 85, 92, 93]. Figure 3a and 3b show force *vs*. deformation traces obtained for single or bilayer graphene and $MoS_2$ layers suspended over circular holes. The experimental traces can be accurately reproduced by expression (2) with $E$ and $T$ as fit parameters, neglecting the contribution of the bending rigidity term.

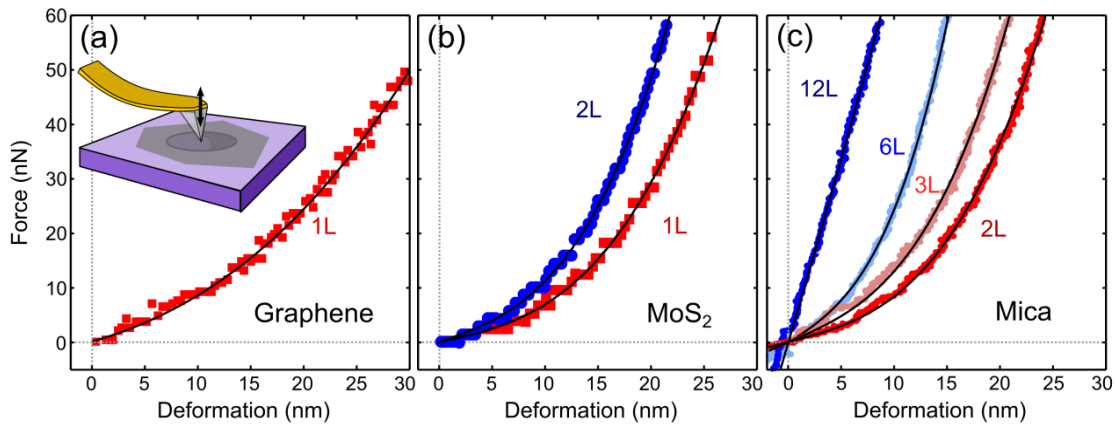

**Figure 3. Central indentation experiments on different 2D materials.** Force *vs*. deformation traces acquired at the center of freely suspended graphene (a), $MoS_2$ (b) and mica (c) drums (adapted from [6], [78] and [63], respectively). The inset in (a) is a schematic diagram of the central indentation experiment carried out with the tip of an atomic force microscope. The solid black lines are the result of fitting the experimental data to Eq. (2) using $E$ and $T$ as fitting parameters: $E_{graphene} \sim 1$ TPa, $T_{graphene} \sim 0.06$ N/m; $E_{MoS2} \sim 270$ GPa, $T_{MoS2} \sim 0.05$ N/m and $E_{mica} \sim 200$ GPa, $T_{mica} \sim 0.14$ N/m. The thickness values employed for the fitting to Eq. (2) correspond interlayer spacing (0.335 nm, 0.65 nm and 1.0 nm for graphene, $MoS_2$ and mica respectively) times the number of layers.

For thick flakes, on the other hand, the bending rigidity can become much larger than the other terms (due to its $t^3$ dependence) and dominate the *F vs. δ* traces which would remain linear even for very large deformations [54, 57, 58, 63, 65]. Therefore, it is expected a non-linear to linear transition of the *F vs. δ* traces as the thickness of the suspended flakes increases. While thinner flakes behave as membranes (tension dominated, negligible bending rigidity), very thick flakes show a plate-like behavior (bending rigidity dominated, negligible tension) as can be seen from the change in the *F vs. δ* traces from non-linear (membrane-like) to linear (plate-like). For intermediate thicknesses, the mechanical behavior of the flakes can only be described by a combination of membrane and plate mechanical behaviors. Figure 3c shows an example of *F vs. δ* traces acquired at the center of mica flakes, from two to 12 layers thick, freely suspended over 1.1 μm diameter holes. The thinner flakes (bilayer to six layers) show marked non-linear traces that can be fitted to Eq. (2) to extract the Young's modulus and the pre-tension. For the thicker flake, on the other hand, the bending rigidity is so large that the *F vs. δ* trace remains linear.

Apart from the elastic properties of 2D materials, the central indentation experiments can be used to study the breaking strength of the freely suspended layers [6, 63, 66, 78, 79, 83, 85]. The intrinsic strength of freely suspended layers is measured by acquiring *F vs. δ* traces with increasingly large force loads until reaching the breaking point. One can estimate the corresponding breaking stress ($\sigma_{max}$) of the layers by using the expression for the indentation of an elastic membrane by a spherical indenter [94, 95]

$$\sigma_{max} = \sqrt{\frac{F_{max}\,E}{4\pi\,r_{tip}\,t}} \qquad , \qquad [3]$$

where $r_{tip}$ is the curvature radius of the AFM tip used.

The breaking strength for exfoliated graphene (130 GPa) [6], CVD grown graphene (~100 GPa) [79, 83, 85], exfoliated $MoS_2$ (10-25 GPa) [66, 78] and exfoliated mica (4-9 GPa) [63] has been recently reported (See Table 1). Despite the different electrical and optical properties of these materials, all of them showed a very large breaking stress, especially if compared to conventional 3D non-layered materials. This property makes 2D materials prospective candidates to be employed in flexible electronics [96-101]. For instance, single-layer graphene and $MoS_2$ have shown very large breaking stress [6, 78], approaching the theoretical limit predicted by Griffith for ideal brittle materials in which the fracture point is dominated by the intrinsic strength of its atomic bonds and not by the presence of defects [102]. While the ideal breaking stress value is expected to be one ninth of the Young's modulus, for graphene and single layer $MoS_2$ it reaches ~1/8. This almost ideal behavior is attributed to a low density of defects on the fabricated devices, probably due to the high crystallinity of the 2D materials in combination to their reduced dimensions.

### 4.2. Spring constant scaling

As pointed out in the previous sub-section, the analysis of the linear term of the *F vs. δ* traces does not allow to unambiguously determine *E* and *T*. Nevertheless, according to expressions (1) and (2) the bending rigidity term depends on the layer thickness while the pretension term does not. Thus, one can determine the Young's modulus and pre-tension of 2D materials by measuring the linear term of the *F vs. δ* traces (*i.e.* the effective spring constant, $k_{eff}$) acquired at the center of suspended flakes with different thicknesses. In fact, according to expressions (1) and (2) the effective spring constant depends on the sample geometry as:

$$k_{\text{eff,doubly-clamped}} = \frac{30.78\,W\,t^3}{L^3}\,E + \frac{12.32}{L}\,T$$

$$k_{\text{eff,circular-drum}} = \frac{4\,\pi\,t^3}{3(1-\nu^2)\,R^2}\,E + \pi\,T \qquad\qquad [4]$$

The typical values for effective spring constants are in the order of ~1.6 Nm$^{-1}$ for graphene, ~0.2 Nm$^{-1}$ for MoS$_2$ and ~0.3 Nm$^{-1}$ for mica (all calculated for circular drums, 1 μm in diameter, using the typical $E$ and $T$ values). This method is usually employed for flakes with thickness in the range from 1 to 15 layers as no strong thickness dependence of the $E$ and $T$ parameters has been observed in this range.

Figure 4a shows $k_{\text{eff}}$ measured for doubly-clamped few-layer graphene flakes with different thicknesses and widths [54]. The Young's modulus can be determined from the slope of the $k_{\text{eff}}$ *vs*. ($W \cdot t^3 / L$) relationship. The pre-tension can be determined from the intercept with the vertical axis. The line in Figure 4a shows the fit of the experimental data to expression (4) to determine the Young's modulus ($E$ = 500 GPa) and the pre-tension ($T$ = 1 N/m) of few-layer graphene flakes.

(a)

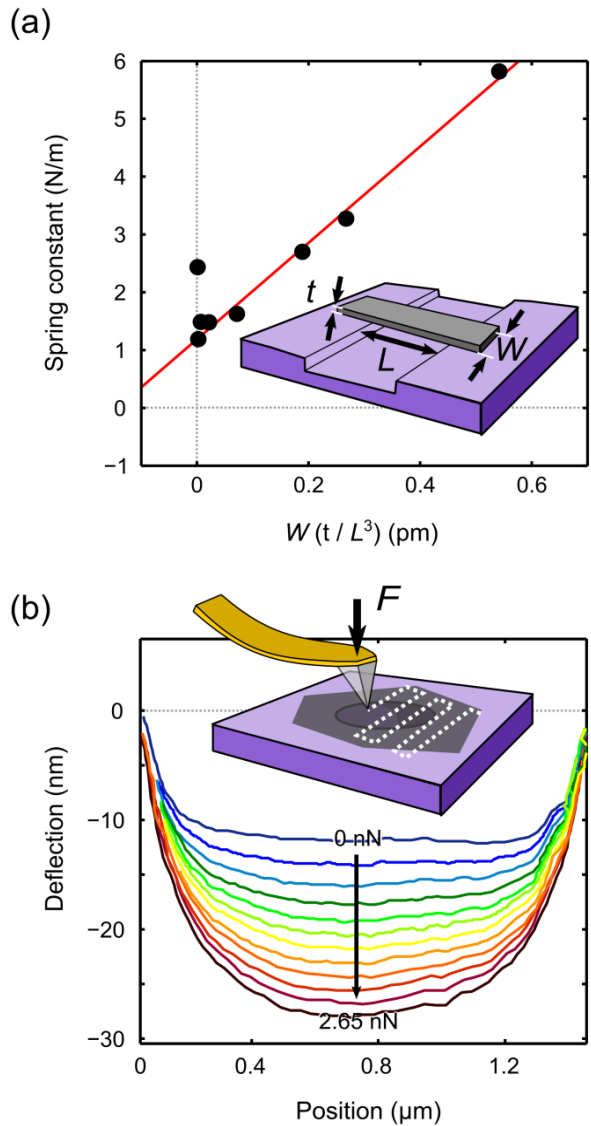

(b)

**Figure 4. Spring constant scaling and constant force maps.** (a) Geometry dependence of the spring constant of graphene-based doubly-clamped beams. A fit to expression (4) is employed to determine the Young's modulus and pre-tension of the devices (adapted from [54]). (b) Topographic line profiles acquired along a graphene oxide flake suspended over a circular hole, acquired at increasingly large force loads (adapted from [76]).

### 4.3. Constant force maps

Alternatively to the previous methods that rely on point indentation measurements, one can also determine the mechanical properties of 2D materials by mapping the deformation (or the compliance) of suspended flakes while an AFM tip is scanned over the whole suspended area at constant force. Figure 4b shows an example of topographic line profiles acquired along an atomically thin graphene oxide flake, suspended over a 1.3 μm diameter hole, at increasing tip-sample forces [76]. Fitting the obtained topographic line profiles to the results of finite element simulations one can determine the Young's modulus and pre-tension of the graphene oxide flake ($E = 207 \pm 23$ GPa and $T = 0.03$-$0.05$ N/m). More details on this technique can be found in the references [76, 86]. A variation of this method consists on making a grid over the suspended hole and acquiring an indentation trace at each point of the grid [57, 65]. The compliance of the flake is then determined at each position and the Young's modulus and pre-tension are obtained by fitting the compliance maps to a continuum mechanics model, see reference [57] for more details.

### 4.4. Electrostatic deflection

The methods described in the previous sub-sections rely on the use of an AFM to apply the force load and to measure the subsequent deformation. Electrostatic force can be also employed to apply accurate force loads to freely suspended flakes [103]. This requires electrical contacts and the flake material should be conducting. While the AFM is employed to apply localized loads, an electrostatic force generates a distributed force profile on the flake. For a circular drum-like suspended flake, with a parallel plate model for the capacitance, the electrostatic force depends on the applied voltage between the flake (connected to a source electrode) and the bottom gate electrode (see Figure 5a and the inset in Figure 5b) as

$$F = \frac{\varepsilon_0 \varepsilon_r A V^2}{2(g-\delta)^2} \quad , \qquad [5]$$

where $\varepsilon_0$ is the vacuum permittivity, $\varepsilon_r$ the relative permittivity of the medium between the flake and the backgate, $A$ is the drum area, $g$ is the drum-backgate distance and $\delta$ is the electrostatically induced deformation. The maximum deformation of the flake for a certain electrostatic force can be calculated from the expression derived in Ref. [59]:

$$\delta = \frac{F\,R^4}{64\,D\,A} + \frac{1}{1+\frac{0.4418}{1-v^2}\frac{\delta^2}{t^2}} \qquad , \qquad [6]$$

where $D$ is the bending rigidity $D = E \cdot t^3 / [12\,(1\text{-}v^2)]$ of the drum. For small deformations ($\delta \ll$ $t$) the second term in expression (6) becomes negligible. Note, that Expression (6) is strictly valid for circular plates with negligible initial pre-tension (details on the derivation can be found in Ref. [59]). Therefore, the method presented in Ref. [59] is strictly valid for thick multilayered flakes where the pre-tension is negligible in comparison to the bending rigidity. For relatively thin flakes, this method should be corrected to include the effect of the initial pre-tension in order to avoid the overestimation of the Young's modulus.

Figure 5b shows two topographic line profiles measured by AFM along a few-layer graphene flake suspended over a 3.8 µm hole and connected to an electrode. When a voltage is applied between the flake and the backgate, the flake feels an electrostatic force and it deflects towards the gate [59]. The Young's modulus can be determined by analyzing the maximum deflection of the flake, when different voltages are applied, in combination with expressions (5) and (6). The value of Young's modulus obtained by this method is compatible to those obtained by AFM-based indentation methods described above (see a comparison of these values in Table 1). An interesting outcome of this technique is that one can directly visualize the deformation induced by electrostatic forces on a freely suspended flake. Therefore, this technique can be used to calibrate the amount of strain obtained with a certain applied voltage. Note, that in mechanical resonators (see Section 6) the frequency of freely suspended flakes is typically tuned by applying an electrostatic force.

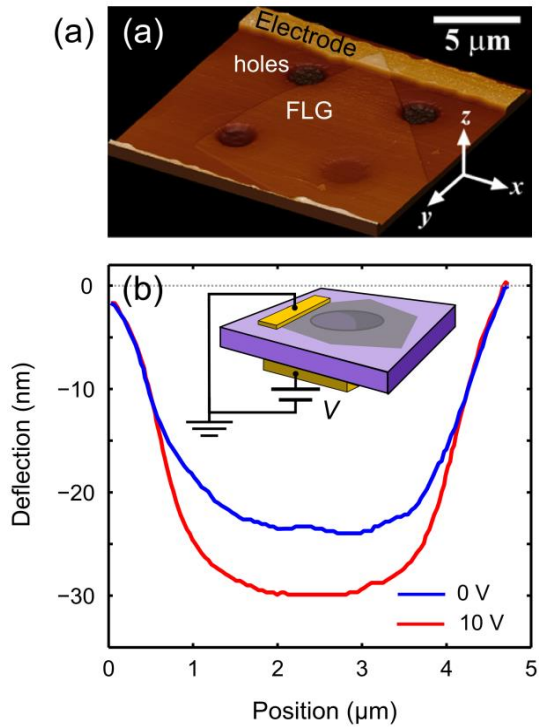

**Figure 5. Electrostatic deflection of 2D materials.** (a) AFM topography image of a freely suspended few-layer graphene flake connected to an electrode. (b) topographic line profiles acquired across the suspended region of the flake at two different applied voltages. The inset in (b) shows an schematic of the electrical connection employed to electrostatically deflect the suspended drum. (Adapted from [59]).

## 4.5. Pressurized blisters

Bunch *et al.* demonstrated that freely suspended graphene membranes are impermeable to standard gases [55]. Therefore, when graphene is transferred onto a substrate with holes (similar to the sample shown in Figure 5a) it forms a microchamber. As the gases cannot diffuse through the graphene membrane, a change in the external pressure leads to the formation of a graphene blister. This is an alternative method to apply a distributed force load. By studying the deflection of pressurized graphene (and few-layer graphene) blisters, Bunch and co-workers determined the Young's modulus of graphene and found it in very good agreement with the 1 TPa value determined by central indentation measurements [67].

## 5.  Comparison of static mechanical properties of different 2D materials

Freely-suspended graphene samples (fabricated by mechanical or liquid-phase exfoliation, CVD based methods or epitaxial growth) have been extensively studied by the characterization methods described in the previous section. The mechanical properties of other 2D materials (like hexagonal boron nitride, mica, $MoS_2$, etc.) have been also recently characterized using similar approaches. Table 1 summarizes the experimental results on mechanical properties of 2D crystals reported in the literature. Information about the employed sample fabrication and characterization methods, the sample geometry and experimental conditions is displayed in the table to facilitate the comparison between different experiments.

In general, freely suspended 2D layered materials show large values of Young's modulus (in the order of 100 to 300 GPa) and low values of pre-tension (see Section 4.1) which makes these materials to be in a different mechanical regime than conventional silicon nitride nanomembranes (10 to 200 nm in thickness) whose mechanics is completely tension-dominated. In the case of graphene, the reported Young's modulus value reaches 1 TPa which makes graphene to be among one of the stiffest materials. The reported values of maximum stress before rupture also show that 2D materials can sustain very large strains without breaking, triggering the interest on strain engineering of their electrical and optical properties.

## 6.  Dynamics of suspended 2D materials: mechanical resonators

In recent years, the community working on nano- electromechancial and optomechanical systems has started to explore the dynamics of freely suspended layered 2D materials. In contrast to the experiments detailed in the previous sections, where the deformation of a suspended 2D crystal upon static force load was studied, in dynamical experiments the oscillation amplitude of suspended flakes subjected to a time varying actuation (resonant or near-resonant) is studied. One can distinguish the mechanical resonances of freely-suspended flakes as peaks of large oscillation amplitude, occurring right when the frequency of the actuation signal is swept across the mechanical resonator resonance frequency. The resonance frequency of the mechanical resonators depends on their geometry and on physical properties of the resonator material (e.g.: pre-tension, mass density and Young's modulus). Therefore the study of the resonance frequency of mechanical resonators with different geometries can be exploited to determine intrinsic mechanical properties of a 2D material, complementing the static approaches described in the previous sections. The width of the resonance peaks is related to the quality factor $Q$, which also gives information about the damping processes in the mechanical resonator.

### 6.1. Actuation and read-out schemes

Transduction schemes translate electrical signals to mechanical displacements (and vice versa) and are crucial for proper resonator operation as one needs to actuate the resonator and to read-out its oscillation. Among the possible actuation mechanisms to drive mechanical resonators based on 2D materials, electrical driving is the most common. Similarly to what was described in the section devoted to the electrostatic deflection, when a (conducting) 2D material is electrically connected to a backgate electrode it forms a capacitor and the electrostatic force between the plates can be used to drive the resonator [8, 12, 53, 56, 60, 71, 72, 77, 80, 82, 84, 87, 89]. This method can be easily implemented in vacuum and cryogenic conditions but it is limited to conductive materials. Another commonly employed driving scheme consists of using a modulated laser signal, focused on the mechanical resonators [9, 11, 55, 69, 74, 77, 81]. The difference in thermal expansion coefficients between the flake and the substrate generates a time-varying strain at the laser modulation frequency that drives the resonator. This driving method can be employed with insulating materials as well, but its application is mostly limited to room temperature measurements (as it is more difficult to implement it in a cryogenic environment than the electrical driving scheme). Also note, that mechanical resonators oscillate, even without any driving force, due to thermal fluctuations. Some experiments probe the mechanical motion of 2D materials driven only by these oscillations [10, 53, 88]. Nonetheless, thermally-induced displacements of the mechanical resonators are usually very small (on the order of ~1pm at room temperature, depending on the exact resonator geometry) making its detection challenging with conventional transduction schemes.

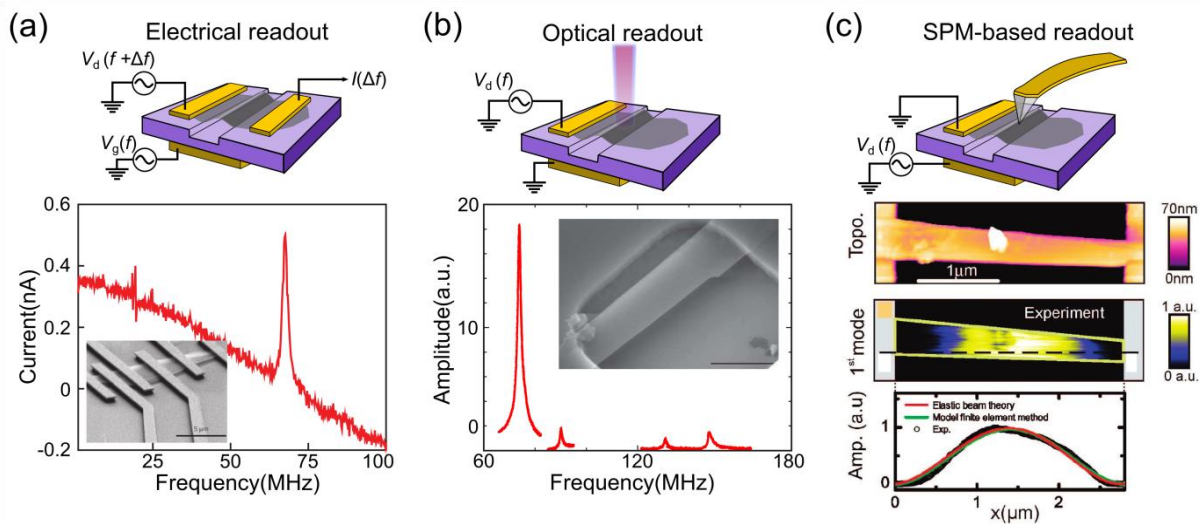

**Figure 6. Oscillation amplitude read-out in mechanical resonators based on 2D materials.** Examples of detection of mechanical resonances by: (a) electrical read-out (adapted from [8]), (b) optical interferometry (adapted from [53]) and (c) scanning probe microscopy (adapted from [56]).

Once the resonator is driven, its oscillation amplitude can be detected by electrical, optical or scanning probe microscopy based methods. The electrical read-out schemes exploit the capacitive coupling between the suspended (conductive) layer and a backgate electrode [8, 12, 60, 71, 72, 77, 80, 82, 84, 87-89]. A detailed description of various electrical-based read-out schemes can be found in Ref. [7]. Similarly to the electrical actuation, electrical read-out (although limited to conductive samples) can be implemented in vacuum and cryogenic environments [8, 12, 71, 72, 77, 80, 82, 87-89].

The optical read-out, on the other hand, can be used to measure insulating materials as well. Although it has been mainly employed in room temperature experiments, it has potential to be implemented in cryogenic conditions. The optical method exploits the change in reflectivity produced by the displacement of the 2D material. In order to detect this change in reflectivity, the studied 2D material is deposited over a hole/trench forming an optical cavity between the flake and the substrate where the 2D layer acts as a semi-transparent mirror [9-11, 53, 69, 74, 77, 81]. The optical path in the cavity, and thus the phase difference between the incoming and reflected light beams, depends on the deflection of the resonator and strongly modifies the overall reflectivity. This facilitates the detection of the oscillation amplitude. A variation of the optical-based read-out, developed in Ref. [104], consists of detecting the motion of few-layer graphene cantilevers by Raman spectroscopy and Fizeau interferometry.

The motion of nanomechanical resonators has been also detected by a scanning probe microscopy based method, allowing even to map the eigenmode shape with ~100 nm resolution [56]. Its implementation is nonetheless more complicated than the previous approaches, hampering its usability in vacuum or cryogenic conditions.

## 6.2. Dimensions scaling of the dynamics of 2D mechanical resonators

Barton *et al*. [11] systematically studied the radius dependence of the resonance frequency of single-layer graphene circular drum resonators. The resonance frequency was found to follow the continuum mechanics model for a circular membrane under initial pre-tension [105]

$$f_{\text{membrane}} = \frac{2.4048}{2\pi R} \sqrt{\frac{T}{\rho\, t}} \qquad , \qquad [7]$$

where $\rho$ ($\rho = 2200$ kg/m$^3$ for graphene) is the 3D mass density. Note, that expression (7) does not account for the contribution of the bending rigidity as it is strictly valid for membranes whose bending rigidity is negligible in comparison to their pre-tension. In the membrane-limit, the resonance frequency is expected to depend on the geometry of the flake as $f \propto 1/(R \cdot t^{1/2})$, according to Eq. (7). Figure 8 shows the resonance frequency measured for several graphene circular resonators with different radius. The solid line in Figure 8 is a fit of the experimental data to expression (7). The membrane-like mechanical behaviour of these graphene circular resonators is also evident from the study of higher eigenmodes. The inset in Figure 8 shows the measured resonance spectra for one graphene resonators, includig several high order eigenmodes. The blue arrows indicate the expected resonance frequencies for a circular membrane under an initial pre-tension.

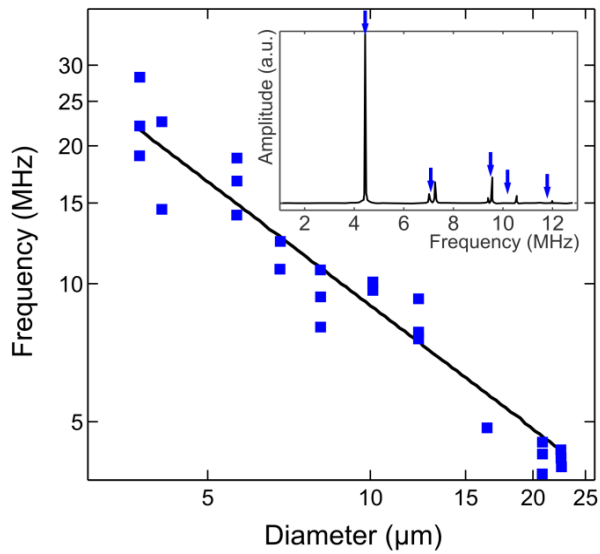

**Figure 7. Membrane-like behavior of graphene-based resonators.** Diameter dependence of the resonance frequency of graphene circular resonators. The experimental data can be fitted to a continuum mechanical model for a circular membrane (neglecting the bending rigidity). (Inset) Fundamental and the first four higher eigenmodes of a circular graphene drum, the arrows indicate the expected positions for the higher eigenmodes of a circular membrane. (Adapted from [11]).

For thicker layers, the bending rigidity becomes more and more important (due to its $t^3$ dependence) and eventually it dominates the dynamics of the mechanical resonator. In the extreme case of a plate-like circular resonator, with negligible initial pre-tension, the frequency is given by [105]

$$f_{\text{plate}} = \frac{10.21}{4\pi}\sqrt{\frac{E}{3\,\rho\,(1-\nu^2)}}\,\frac{t}{R^2} \quad . \qquad [8]$$

In the plate-limit, the resonance frequency is expected to depend on the geometry of the flake as $f \propto t/R^2$, according to Eq. (8). This dependence is significantly different than that expected in the membrane limit ($f \propto t^{1/2}/R$). Figure 8a shows the resonance frequency measured for $MoS_2$ circular resonators (6 μm in diameter) with different thicknesses, ranging from 15 to 100 layers [10]. The solid lines present the resonance frequency for different thicknesses, calculated for the membrane and plate limiting cases. From Figure 8a, it is evident that flakes in the thickness range above 15 layers cannot be modelled by the membrane model, expression (7), but they follow the dynamics expected for plates, expression (8).

For certain thickness, the bending rigidity and initial pre-tension are comparable and the mechanical resonators dynamics is thus in the membrane-to-plate crossover regime. The resonance frequency of the mechanical resonators can be then approximated by

$$f \approx \sqrt{f_{\mathrm{membrane}}^2 + f_{\mathrm{plate}}^2} \quad , \qquad [9]$$

where $f_{\mathrm{membrane}}$ and $f_{\mathrm{plate}}$ are given by Eq. (7) and (8) respectively. The inset in Figure 8a shows the measured resonance frequency for $MoS_2$ circular resonators (3 μm in diameter) with thicknesses ranging from single layer to ~100 layers [9]. The *f vs. t* relationship, calculated for the two limiting cases (pre-tension dominated and bending rigidity dominated), has been plotted for comparison. The Young's modulus and pre-tension, determined from central indentation measurements, have been employed for the calculation using Eqs. (7) and (8). The frequency for pre-tension dominated resonators (membrane limit) decreases as a function of the thickness. While thin flakes (1 to 5 layers) follow this trend, thicker flakes strongly deviate from it. In fact, the resonance frequency of thicker flakes increases with the thickness, as expected from bending rigidity dominated resonators (plate limit). Interestingly, mechanical resonators with thickness in the range of 4 to 10 layers are in a crossover regime where both terms in Eq. (9) are needed to accurately describe their dynamics. This thickness dependent membrane-like or plate-like mechanical behavior is also observed in static central indentation measurements on freely-suspended $MoS_2$ [64, 65].

Further evidence for the membrane-to-plate-like crossover can be deduced from an analysis of high-order eigenmodes. Figure 8b shows an example of the resonance spectra measured for two $MoS_2$ drums with different thicknesses (single layer in the top panel; nine layers in the bottom panel). The blue arrows indicate the expected frequencies of the higher-order modes for the membrane case (top panel) and plate case (bottom panel). Note, that for the membrane case one expects the second eigenmode to be at 1.56 times the fundamental mode frequency, while for the plate case the second eigenmode occurs at twice the fundamental frequency (see Ref. [105] for a detailed discussion about the higher eigenmodes in these two limiting cases). Therefore, the analysis of the second eigenmode can be used to determine whether the mechanical resonator behaves as a tensed membrane or as an elastic plate without tension.

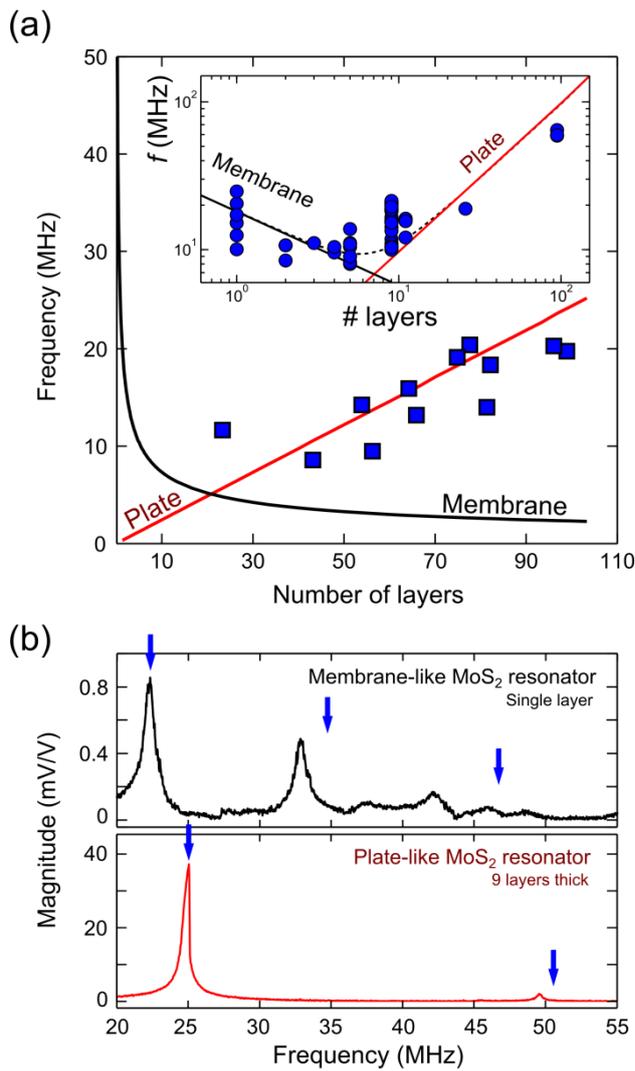

**(a)**

**(b)**

**Figure 8. Membrane–to-plate crossover.** (a) Thickness dependence of the resonance frequency of thick multilayered $MoS_2$ circular resonators (6 μm in diameter) (adapted from [10]). The calculated thickness dependence for circular membranes and circular plates has been included for comparison ($E$ = 270 GPa and $T$ = 0.1 N/m were used in the calculation). (Inset) Continuous crossover from membrane to plate mechanical behavior observed for thinner $MoS_2$ resonators (3 μm in diameter), adapted from [9]. (b) Comparison between the higher eigenmodes measured for $MoS_2$ mechanical resonators in the membrane (top) and plate (bottom) limits. The blue arrows indicate the expected resonance frequencies for the higher modes in the membrane (top) and plate (bottom) cases. (Adapted from [9]).

## 7. Comparison of mechanical resonators based on different 2D materials

In the past years, different 2D materials have been employed to fabricate mechanical resonators and to measure their dynamical response. Graphene and graphene oxide are the most studied 2D materials so far. In fact, graphene-based mechanical resonators have been fabricated with different geometries (although doubly-clamped beams and drumheads are the most common geometries there are also recent works on graphene-based cantilevers [104, 106]) and studied under various environmental conditions. Reports on nanomechanical systems based on exfoliated [12, 53, 55, 56, 60, 72, 80, 87-89], epitaxially grown [69] and CVD graphene [11, 77, 82, 84], graphene oxide [74, 81], exfoliated $MoS_2$ [9, 10, 107, 108] and $NbSe_2$ [71] can be found in the literature. Figure 9 shows some examples of mechanical responses of resonators based on different 2D materials. Note that despite the large variety of 2D materials that can be isolated, the reports on their resonating behaviour are still limited to the materials displayed in Figure 9.

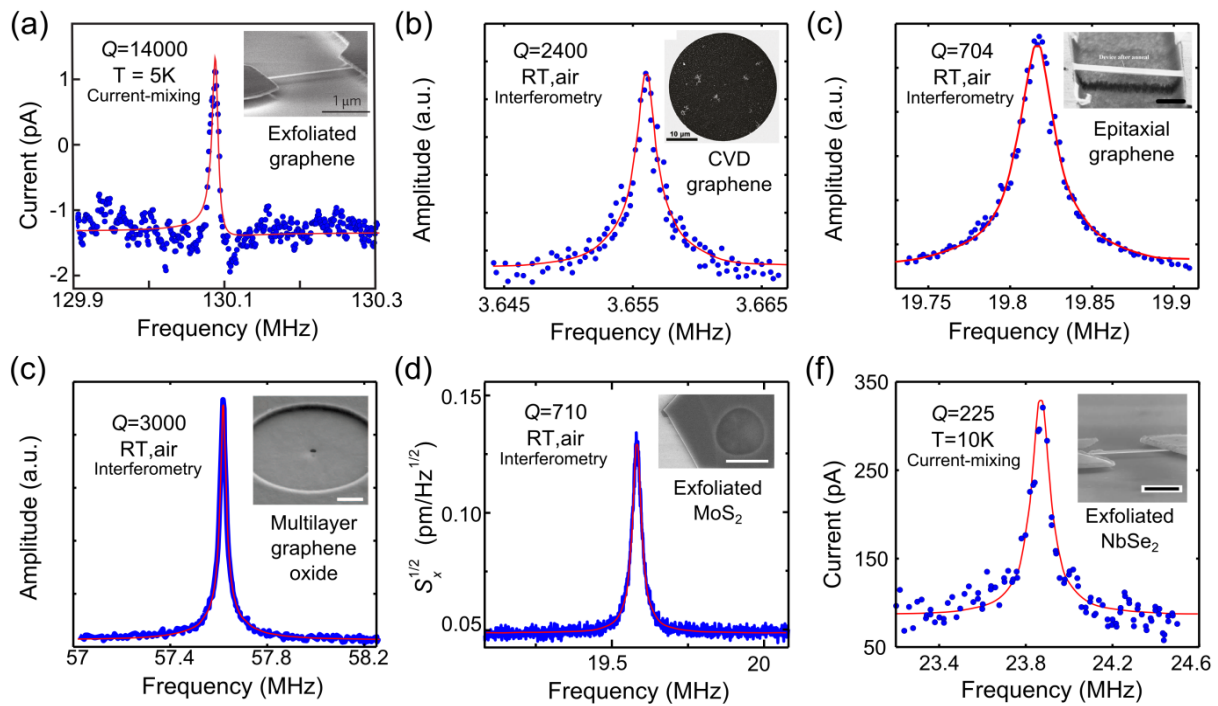

**Figure 9. Mechanical resonance measured for several resonators based on different 2D materials**. Resonance spectra of mechanical resonators based on: (a) exfoliated graphene (adapted from [8]), (b) CVD grown graphene (adapted from [11]), epitaxial graphene (adapted from [69]), (d) graphene oxide (adapted from [74]), (e) exfoliated $MoS_2$ (adapted from [10]) and (f) $NbSe_2$ (adapted from [71]). The quality factor and measurement conditions are displayed in each plot.

Graphene-based mechanical resonators typically show resonance frequencies in the range of 0.4 MHz to 200 MHz (depending on the exact sample geometry). This makes graphene mechanical resonators very suitable for very high frequency (VHF) electronic applications such as filters or oscillators [82]. The reported quality factor of graphene nanomechanical resonators spans from 2 to $2 \cdot 10^5$ as it is strongly dependent on the fabrication and measurement conditions. The largest quality factor reported for 2D materials has been achieved for dry transferred multilayer graphene ($Q \sim 220000$) [90], measured at low temperatures (<50 mK) by an electrical based actuation/read-out scheme (coupling the motion of the graphene-based resonator to a superconducting microwave cavity) [72]. Note, that although it is well-known that the quality factor substantially increases at cryogenic temperatures the origin of this increase is still subject of debate as many damping mechanisms could be responsible of the reduced quality factors at room temperature.

At room temperature, other mechanical resonators based on alternative 2D materials, beyond graphene, show resonance frequencies within the same range as graphene. Their quality factor, measured in vacuum, is also rather similar to that of graphene (measured at similar conditions). Low temperature measurements on 2D materials different from graphene would be necessary to address whether the high quality factor observed in graphene is a general property of 2D crystals or not. However, experiments on mechanical resonators based on other 2D materials in cryogenic environment are still very scarce [71].

Table 2 presents a summary of the experimental results reported for mechanical resonators based on 2D materials. The device fabrication method and geometry as well as the actuation/read-out scheme and experimental conditions are included to facilitate the comparison.

## 8   Hybrid systems with atomically thin layered materials

The low mass density and high crystallinity of atomically thin materials also give them a unique edge to form mechanically active element for fabricating hybrid systems. Low pre-stress, high frequency, and high quality factors leads to large quantum zero point fluctuations and hence larger coupling strengths. Further their interesting optical and electronic properties can host new types of hybrid couplings with optical and microwave photons. The most explored atomically thin material for hybrid devices so far is graphene. Photothermal and radiation pressure couplings have been explored experimentally and have demonstrated strong backaction at room temperature [91] and at cryogenic temperatures [87-89]. There are also proposals based to coupling scheme mediated by the vacuum forces using these materials [109].

Photothermal hybrid coupling of graphene resonator to an optical cavity was demonstrated by Barton *et al.* [91]. The essential idea behind the photothermal coupling is that when laser light is shined on a suspended graphene flakes, it modifies the tension in the flake (heating due to light absorption) which leads to the change in its mean position. By combining it with an optical cavity such a change in the mean position places the resonator at a different intensity of the cavity light field leading to a retardation and hence a backaction.

Owing to graphene's strong and uniform absorption over a wide range of wavelengths ($\pi\alpha \sim$ 2.3%) [110], Barton *et al.* couples a 10 μm × 10 μm CVD grown single layer graphene flake to a low finesse (due to low reflection coefficient of graphene) Fabry-Parrot optical cavity formed by a platinum gate and graphene itself as shown in Figure 10(a). The presence of a conducting backplane further allows one to apply a DC gate voltage and tune the tension in the graphene flake (frequency tuning of more than 100%). As the mean displacement of the flake can be controlled by the DC gate voltage, the sign of the damping due to photothermal backaction can be controlled too. Graphene motion can be damped as well as can be amplified by controlling the laser power and gate voltage. Furthermore, with sufficiently high laser powers, it is possible to make the total mechanical dissipation negative, which leads to self-oscillations of the graphene membrane driven by the laser light, demonstrating a significant backaction due to photothermal effect. The coupling to a low finesse cavity further allows one to achieve a displacement sensitivity of 600 fm/√Hz at room temperature. However, as the photothermal effect relies on the absorption of the light by graphene and hence heating, it has to compete with cooling produced by the backaction to take these systems towards the quantum regime. Alternatively, one can also use the radiation pressure induced backaction to improve the cooling methods for reaching the quantum ground state, however, this would require a high finesse (low loss) cavity [111], which is not possible due to the low reflection coefficient and large optical absorption of graphene [110].

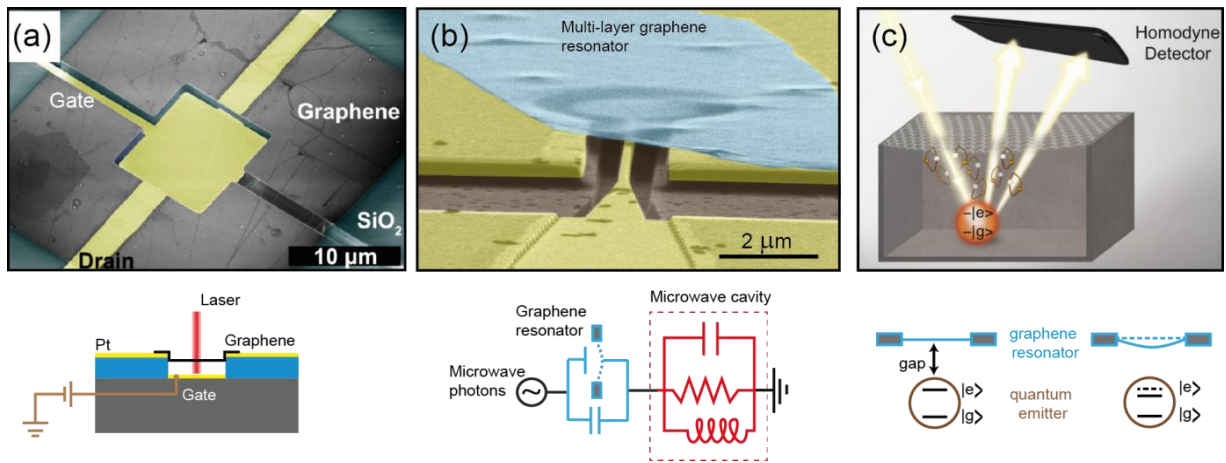

**Figure 10. Hybrid systems with atomically thin crystals:** (a) False color scanning electron microscope (SEM) image of a suspended graphene resonator forming a low finesse optical cavity with platinum gate electrode. By shinning a laser, photothermal coupling can produce a significant backaction controllable by laser power and gate voltage, illustrated in the schematic (Adapted from Ref. [91]) (b) False color SEM image of a multi-layer graphene resonator coupled to a high quality factor superconducting microwave cavity. This coupling allows one to apply radiation pressure on the graphene mechanical resonator using the sideband techniques (Adapted from Ref. [87]). (c) Schematic of a graphene resonator coupled to a quantum emitter (two level system). The mechanical motion of the resonator can be coupled to the quantum emitter via the vacuum forces. This can be detected by measuring the phase of the scattered light from the quantum emitter. (Adapted from Ref. [109]).

High quality factor optical cavities with graphene are possible in the microwave domain using superconductors [87-89]. Previously, superconducting hybrid systems have been successfully used to cool mechanical resonators to their quantum ground state [112, 113]. The large electrical conductivity of graphene allows one to capacitively couple it to superconducting circuits with minimal losses. However, combing the superconducting cavity fabrication technology with graphene's scotch tape technology and yet retaining their pristine properties is challenging. Following a dry transfer scheme [90], Singh *et al.* [87] have been able to form a optomechanical system where a multi-layer graphene mechanical resonator is capacitively coupled to a high quality factor superconducting cavity as shown in Figure 10(b).

This coupling scheme allows measuring the thermo-mechanical motion of graphene resonator at temperature as low as 96 mK with displacement sensitivity down to 17 fm /√Hz. The dry transfer method and nearly circular geometry (low clamping losses) allow measuring large mechanical quality factor as high as 220,000, thus far the largest reported for an exfoliated crystal. The large quality factor of the superconducting microwave cavity allows producing significant radiation pressure backaction on the graphene resonator reflecting in the optomechanical effects such as the onset of normal mode splitting and mechanical microwave amplification. In a similar experiment, Song *et al.* [88] have also demonstrated significant radiation pressure back action on a bilayer graphene resonator by capacitively coupling it to a lumped inductor and hence forming a microwave cavity. With this approach, using the sideband cooling technique they demonstrate a lowest phonon occupation of 40.

Taking the ideas from optomechanics further, there is a proposal of coupling mechanics to quantum emitters (two level system) mediated by the strong vacuum interactions [109] as shown in the schematic in Figure 10(c). The movement of the graphene sheet leads to the modification of the vacuum force (Casimir force) experienced by the emitter, leading to the modification of its frequency. Emitter's frequency can be monitored by the measuring the phase shift of the scattered field. By placing a quantum emitter as close as 30 nm from the graphene resonator, coupling strength as high as 1 GHz/nm are predicted, enabling squeezing of the graphene motion on shorter time scales than the mechanical period. This theoretical proposal is also attractive due to the single emitter nature of the optical cavity, which should enable creation and detection of non-classical states of the mechanical motion.

From these initial experiments, the importance of 2D materials for making hybrid crystals is quite evident. It suggests that apart from graphene, few-layer thick crystals with higher reflectivity and optical conductivity 2D crystals could be better to improve the performance of these hybrid devices. Being excellent host and high Q mechanical resonators, their electronic properties can provide new routes for hybrid coupling.

# 9  Perspectives

The mechanical properties of graphene and graphene-based mechanical resonators have been extensively studied in the past years. However, little is still known about different 2D materials although the amount of reports on layered materials that can be isolated/exfoliated down to single-layer or few-layers keeps growing [31]. In the past four years, many 2D materials different from graphene have been exploited in nanoelectronic and optoelectronic devices. The use of atomically thin materials different than graphene in nanomechanical systems may open the door to study new physical phenomena as the 2D materials family presents a rich variety of properties. Nonetheless, examples of nanomechanical devices that exploit 2D materials beyond graphene to observe new phenomena are still very scarce. The integration of transition metal chalcogenide 2D materials, with a rich variety of electronic transport phenomena, in mechanical resonators is a very interesting avenue in the field of nanomechanics. One example is a mechanical resonator fabricated from an exfoliated few-layer $NbSe_2$ flake that has been used to study the effect of the charge density wave transition on the mechanical properties [71].

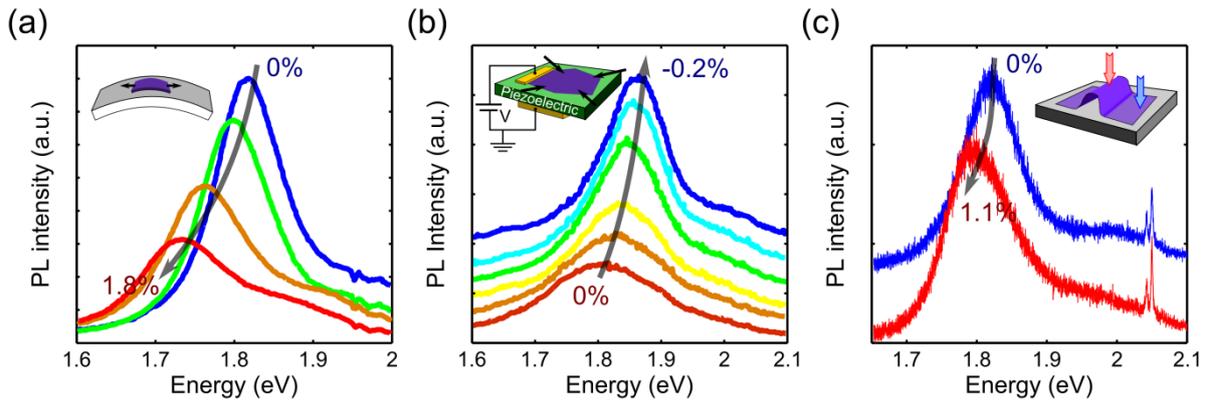

**Figure 11. Strain tunable optoelectronic properties of 2D semiconductors.** (a) to (c) Show photoluminescence spectra acquired for atomically thin MoS$_2$ layers subjected to different strains: uniform uniaxial (a), uniform biaxial (b) and non-uniform uniaxial (c). The photoluminescence peak (related to the direct bandgap transition) is shifted in energy with the applied strain. (The data has been adapted from [114], [115] and [116])

Another aspect of the 2D materials research that is attracting a great deal of attention recently is strain engineering [117-128]. Although most of the strain engineering experiments have been carried out on non-suspended flakes [114, 115, 129], the combination of this strain tunability of the optoelectronic properties with the fabrication of freely suspended mechanical resonator has the potential to yield new read-out schemes for these nanometer scale systems where the motion of the resonator is probed by the shift of the optical absorption or photoluminescence emission due to the strain induced during the oscillation of the 2D material. Atomically thin MoS$_2$, for example, can sustain about 20-50 times larger deformations than silicon before rupture. This fact has triggered the interest on strain tuning the optoelectronic properties of atomically thin materials [114-116, 129-131]. Figure 11 shows photoluminescence spectra acquired for atomically thin MoS$_2$ samples subjected to different strain schemes (uniaxial, biaxial and bending). The photoluminescence peak (which is related to the direct bandgap transition energy) changes with the applied strain, demonstrating that the strain can be effectively used to modify the electronic bandstructure in atomically thin materials.

## 8. Conclusions

Atomically thin two-dimensional materials have opened a new direction in nanomechanics due to their inherent low dimensions and reduced mass. Recently, novel experimental methods have been developed to study the mechanical properties of these nanomaterials and the mechanical properties of 2D materials with very different properties (conductors, semiconductors and insulators) have been recently studied. In this article the different fabrication and characterization methods to study the mechanical properties of 2D materials are described and discussed. A thorough comparison of the mechanical properties of different 2D materials is presented, focusing on the membrane-to-plate like crossover. This article also reviews these recent experimental works on nanomechanical devices, comparing the different results in the literature.

## Acknowledgements

This work was supported by the European Union (FP7) through the program 9 RODIN and the Dutch organization for Fundamental Research on Matter (FOM). A.C-G. acknowledges financial support through the FP7-Marie Curie Project PIEF-GA-2011-300802 ('STRENGTHNANO').

**Table 1.** Summary of the reported static mechanical deformation experiments of freely suspended 2D materials. The sample thickness and geometry as well as the measuring technique and conditions have been included to facilitate the comparison.

| Material | Number of layers | Geometry | $E$ [Gpa] | $\sigma_{max}$ [GPa] | Experimental conditions | Method | Ref. |
|---|---|---|---|---|---|---|---|
| Graphene | 6-20 | Doubly-clamped beam | 500 | -- | RT, ambient | Spring constant scaling | [54] |
| | 1 | Circular drum | 1000 ± 100 | 130 ± 10 | RT, ambient | Central indentation | [6] |
| | 8-100 | Circular drum | 920 | -- | RT, ambient | Compliance maps | [57] |
| | 23-43 | Circular drum | 1000 | -- | RT, ambient | Electrostatic deflection | [59] |
| | 1 | Doubly-clamped beam | 800 | -- | RT, ambient | Central indentation | [61] |
| | 1 | Circular drum | 1120 | -- | RT, ambient | Central indentation | [62] |
| | 2-5 | Circular drum | 3225-3430 | -- | RT, ambient | Central indentation | [67] |
| | 1-5 | Circular drum | ~1000 | -- | RT, < 1.75 MPa | Pressurized blister | |
| | 1 | Doubly-clamped beam | 430 | -- | RT, ambient | Spring constant scaling | [70] |
| | 1 | Circular drum | 800-1100 | -- | RT, ambient | Central indentation | [93] |
| (Reduced) Graphene oxide | 1 | Doubly-clamped beam | 250 ± 150 | -- | RT, ambient | Spring constant scaling | [68] |
| Graphene oxide | 1 | Circular drum | 207.6±23.4 | -- | RT, ambient | Constant force maps | [76] |
| Graphene (CVD) | 1 | Circular drum | 160 | 35 | RT, ambient | Central indentation | [79] |
| | 1 | Circular drum | 1000 ± 50 | 103-118 | RT, ambient | Central indentation | [83] |
| | 1 | Circular drum | -- | 90-94 | RT, ambient | Central indentation | [85] |
| MoS$_2$ | 5-10 | Circular drum | 330 ± 70 | -- | RT, ambient | Central indentation | [64, 65] |
| | 5-25 | Circular drum | 290 ± 80 | -- | RT, ambient | Spring constant scaling | |
| | 8 | Circular drum | 400 ± 30 | -- | RT, ambient | Compliance maps | |
| | 1 | Circular drum | 210 ± 50 | 26.8 ± 5.4 | RT, ambient | Central indentation | [66] |
| | 1 | Circular drum | 270 ± 100 | 16-30 | RT, ambient | Central indentation | [78] |
| | 2 | Circular drum | 200 ± 60 | -- | RT, ambient | Central indentation | |
| Na$_{0.5}$-Fluorohectorice | 12-90 | Doubly-clamped beam | 21 ± 9 | -- | RT, ambient | Spring constant scaling | [58] |
| Mica | 2-8 | Circular drum | 200 ± 30 | 4-9 | RT, ambient | Central indentation | [63] |
| | 2-14 | Circular drum | 170 ± 40 | -- | RT, ambient | Spring constant scaling | |
| hBN (CVD) | 2 | Circular drum | 223 ± 16 | -- | RT, ambient | Central indentation | [75] |
| Vermiculite | >2 | Circular drum | 175 ± 16 | -- | RT, ambient | Constant force maps | [86] |

**Table 2.** Summary of the experimental works on nanomechanical systems based on freely suspended 2D materials. The sample thickness and geometry as well as the actuation/read-out method and experimental conditions have been included to facilitate the comparison.

| Material | # layers | Geometry | $f$ (MHz) | $Q$ factor | Experimental conditions | Driving | Read-out | Ref. |
|---|---|---|---|---|---|---|---|---|
| **Graphene** | 1-143 | Doubly-clamped beam | 10-170 | 20-850 | RT, vacuum | Electrical / Thermal noise | Optical | [53] |
| | 1 | Square drum | 30-90 | 25 | RT, variable pressure | Optical | Optical | [55] |
| | 3-57 | Doubly-clamped beam | 18-57 | 2-30 | RT, ambient | Electrical | SPM-based | [56] |
| | 1 | Doubly-clamped beam | 33-36 | 10000 | 77K, vacuum | Electrical | Electrical | [60] |
| | 1 | Doubly-clamped beam | 30-120 | 125 | RT, vacuum | Electrical | Electrical | [8] |
| | 1 | Doubly-clamped beam | 30-120 | 14000 | 5K, vacuum | Electrical | Electrical | |
| | 1 | Doubly-clamped beam | 55-120 | 500-2500 | 7K, vacuum | Electrical | Electrical | [72] |
| | 1 | Doubly-clamped beam | 150-200 | 100000 | 90mK, vacuum | Electrical | Electrical | [12] |
| | 1 | Doubly-clamped beam | 55-180 | 1400 | 4.2K, vacuum | Electrical | Electrical | [80] |
| | 30 | Circular drum | 36 | 220000 | 14mK, vacuum | Electrical | Electrical | [87] |
| | 2 | Doubly-clamped beam | 23 | >10000 | 22mK, vacuum | Thermal noise | Electrical | [88] |
| | 3-4 | Circular drum | 33-60 | 100000 | 33mK, vacuum | Electrical | Electrical | [89] |
| | ~100 | Cantilever | 0.4-1.2 | 2 | RT, air | Electrical | Optical | [104] |
| | ~100 | Cantilever | 0.4-1.2 | 22 | RT, vacuum | Electrical | Optical | |
| | ~100 | Cantilever | 0.44 | 25 | 15K, vacuum | Electrical | Optical | |
| **Graphene (SiC)** | 1 | Doubly-clamped beam | 3-50 | 97-704 | RT, vacuum | Optical | Optical | [69] |
| | 3 | Doubly-clamped beam | 5.3-7.5 | 50-600 | RT, vacuum | Piezo-diether | Optical | [132] |
| | 3 | Doubly-clamped beam | 5.3-7.5 | 2500 | 10K, vacuum | Piezo-diether | Optical | |
| **Graphene (CVD)** | 1 | Doubly-clamped beam | 3-25 | 20-280 | RT, vacuum | Optical | Optical | [77] |
| | 1 | Doubly-clamped beam | 3-25 | ~10000 | 10K, vacuum | Electrical | Electrical | |
| | 1 | Circular drum | 5-25 | 200-2400 | RT, vacuum | Optical | Optical | [11] |
| | 1 | Circular drum | 47-52 | 55 | RT, vacuum | Electrical | Electrical | [82] |
| | 1 | Circular drum | 45-50 | 60 | RT, vacuum | Electrical | Electrical | [84] |
| | 1 | Square drum | 3.4-18 | ~500 | RT, vacuum | Electrical | Optical | [91] |
| | 1 | Circular drum | 0.72 | 700 | RT, vacuum | Optical | Optical | [133] |
| | 1-3 | Cantilever | 0.1-10 | ~2 | RT, vacuum | Optical | Optical | [106] |
| **(Reduced) Graphene oxide** | 11-28 | Circular drum | 15-60 | 400-3000 | RT, vacuum | Optical | Optical | [74] |
| | 28-114 | Circular drum | 8-40 | 2000-31000 | RT, vacuum | Optical | Optical | [81] |
| **NbSe₂** | 45-80 | Doubly-clamped beam | 20-40 | 215 | 4-70K, vacuum | Electrical | Electrical | [71] |
| **MoS₂** | 10-108 | Circular drum | 8-58 | 40-710 | RT, vacuum | Thermal noise | Optical | [10] |
| | 1-10 | Circular drum | 8-36 | 18-360 | RT, vacuum | Optical | Optical | [9] |
| | 11-85 | Circular drum | 12-65 | 48-445 | RT, vacuum | Optical | Optical | |